\documentclass{article}
\usepackage{amsmath}
\usepackage{graphicx}
\begin{document}
\title{
Ward identities for anisotropic Cooper pairs
}

\author{
O. Narikiyo
\footnote{
Department of Physics, 
Kyushu University, 
Fukuoka 812-8581, 
Japan}
}

\date{
(Aug. 26, 2011)
}

\maketitle
\begin{abstract}
Ward identities for anisotropic Cooper pairs are derived. 
These for nonlocal pairs have the same form 
as those for local pairs 
by employing the pair propagator with the form factor.
\end{abstract}

\vskip 30pt 

\section{Introduction}

Recently I have reported the derivation 
of Ward identities for Cooper pairs~\cite{Nar}. 
There only the local pairs are considered 
so that the derivation is quite simple. 
In this note I will extend it to the case of nonlocal pairs. 
For the sake of clarity 
I will only discuss the case of zero temperature. 
At finite temperature 
we can use the same Ward identity with thermal frequency~\cite{Nar}. 
Following description is based on ref.~\cite{Nar}. 

\section{Ward identities for electrons}

First we derive the Ward identity for electrons. 
While the local description in coordinate space 
is quite simple as shown in ref.~\cite{Nar}, 
we use the description in momentum space 
preparing the discussion for nonlocal Cooper pairs. 

Under the charge-conservation law 
the divergence of the three-point function $\Lambda_\mu^e$ 
is expressed as 
\begin{align}
\sum_{\mu=0}^3 {\partial \over \partial z_\mu} \Lambda_\mu^e(x,y,z) = 
& \langle T 
\{ [j_0^e(z), \psi_\uparrow(x)] \psi_\uparrow^\dag(y) \} \rangle 
\delta(z_0-x_0) \nonumber \\ 
+ 
& \langle T 
\{ \psi_\uparrow(x) [j_0^e(z), \psi_\uparrow^\dag(y)] \} \rangle 
\delta(z_0-y_0). \label{divLamxyz-e} 
\end{align}
Performing the Fourier transform we obtain 
\begin{align}
\sum_{\mu=0}^3 i k_\mu \Lambda_\mu^e(p,p-k) = 
\int d & (x_0-y_0) e^{-ip_0(x_0-y_0)} \int d (z_0-x_0) e^{-ik_0(z_0-x_0)} 
\nonumber \\ 
\times \Bigl(
& \langle T 
\{ [j_{\vec k}^e(z_0), a_{{\vec p}-{\vec k}}(x_0)] a_{\vec p}^\dag(y_0) \} 
\rangle \delta(z_0-x_0) \nonumber \\ 
+ 
& \langle T 
\{ a_{{\vec p}-{\vec k}}(x_0) [j_{\vec k}^e(z_0), a_{\vec p}^\dag(y_0)] \} 
\rangle \delta(z_0-y_0) \Bigr), \label{divLambda-e} 
\end{align}
where we have assumed the translational invariance of the system. 
The electric charge $j_0^e(z)$ is transformed as 
\begin{equation}
j_{\vec k}^e(z_0) = 
\int d{\vec z} e^{-i{\vec k}\cdot{\vec z}} j_0^e(z) 
\equiv j_{\vec k}^e = e 
\sum_{\vec p} 
\Bigl( a_{{\vec p}-{\vec k}}^\dag a_{\vec p} 
     + b_{{\vec p}-{\vec k}}^\dag b_{\vec p} \Bigr). 
\end{equation}
The equal time commutation relations are estimated as 
\begin{equation}
[ j_{\vec k}^e, a_{{\vec p}-{\vec k}} ] = - e a_{\vec p}, \ \ \ \ \ 
[ j_{\vec k}^e, a_{\vec p}^\dag ] = e a_{{\vec p}-{\vec k}}^\dag. 
\end{equation}
Thus the divergence in the momentum space becomes 
\begin{align}
\sum_{\mu=0}^3 i k_\mu \Lambda_\mu^e & (p,p-k) = 
\int d (x_0-y_0) e^{-ip_0(x_0-y_0)} \int d (z_0-x_0) e^{-ik_0(z_0-x_0)} 
\nonumber \\ 
& \times \Bigl(
{ - i e } G_{\vec p}(x_0-y_0) \delta(z_0-x_0) 
+ 
{ i e } G_{{\vec p}-{\vec k}}(x_0-y_0) \delta(z_0-y_0) 
\Bigr), \label{intGp-e}
\end{align}
where 
\begin{equation}
G_{\vec p}(x_0-y_0) = 
- i \langle T \{ a_{\vec p}(x_0) a_{\vec p}^\dag(y_0) \} \rangle. 
\end{equation}
Introducing the Green function with four-momentum 
\begin{equation}
G(p) = 
\int d (x_0-y_0) e^{-ip_0(x_0-y_0)} G_{\vec p}(x_0-y_0), 
\end{equation}
we obtain 
\begin{equation}
\sum_{\mu=0}^3 k_\mu \Lambda_\mu^e(p,p-k)
= - e G(p) + e G(p-k). 
\end{equation}
Then shifting the four-momentum this relation is written into 
\begin{equation}
\sum_{\mu=0}^3 k_\mu \Lambda_\mu^e(p+k,p)
= e G(p) - e G(p+k). 
\end{equation}
In terms of the vertex function $\Gamma_\mu^e$ 
where 
$\Lambda_\mu^e(p+k,p) = i G(p+k) \cdot \Gamma_\mu^e(p+k,p) \cdot i G(p)$, 
we finally obtain 
the Ward identity for electric current vertex 
\begin{equation}
\sum_{\mu=0}^3 k_\mu \Gamma_\mu^e(p+k,p) = 
e G^{-1}(p) - e G^{-1}(p+k). \label{WI-e} 
\end{equation}

Under the energy-conservation law 
the divergence of the three-point function $\Lambda_\mu^Q$ 
is expressed as 
\begin{align}
\sum_{\mu=0}^3 i k_\mu \Lambda_\mu^Q(p,p-k) = 
\int d & (x_0-y_0) e^{-ip_0(x_0-y_0)} \int d (z_0-x_0) e^{-ik_0(z_0-x_0)} 
\nonumber \\ 
\times \Bigl(
& \langle T 
\{ [j_{\vec k}^Q(x_0), a_{{\vec p}-{\vec k}}(x_0)] a_{\vec p}^\dag(y_0) \} 
\rangle \delta(z_0-x_0) \nonumber \\ 
+ 
& \langle T 
\{ a_{{\vec p}-{\vec k}}(x_0) [j_{\vec k}^Q(y_0), a_{\vec p}^\dag(y_0)] \} 
\rangle \delta(z_0-y_0) \Bigr). \label{divLambda-Q} 
\end{align}
Using the equal time commutation relations 
\begin{equation}
[ j_{\vec k}^Q, a_{{\vec p}-{\vec k}} ] \Rightarrow 
[ H, a_{\vec p} ], \ \ \ \ \ 
[ j_{\vec k}^Q, a_{\vec p}^\dag ] \Rightarrow 
[ H, a_{{\vec p}-{\vec k}}^\dag ], 
\end{equation}
and the equation of motion 
\begin{equation}
[ H, a_{\vec p}(x_0) ] = - i 
{\partial \over \partial x_0} a_{\vec p}(x_0), \ \ \ \ \ 
[ H, a_{{\vec p}-{\vec k}}^\dag(y_0) ] = - i 
{\partial \over \partial y_0} a_{{\vec p}-{\vec k}}^\dag(y_0), 
\end{equation}
the divergence is written into 
\begin{align}
\sum_{\mu=0}^3 i k_\mu & \Lambda_\mu^Q (p,p-k) = 
\int d (x_0-y_0) e^{-ip_0(x_0-y_0)} \int d (z_0-x_0) e^{-ik_0(z_0-x_0)} 
\nonumber \\ 
& \times \Bigl(
{\partial \over \partial x_0} G_{\vec p}(x_0-y_0) \delta(z_0-x_0) 
+ 
{\partial \over \partial y_0} G_{{\vec p}-{\vec k}}(x_0-y_0) \delta(z_0-y_0) 
\Bigr), \label{intGp-Q} 
\end{align}
Here it should be noted 
that the replacement of the commutation relation ($\Rightarrow$) holds 
in the limit of vanishing external momentum, ${\vec k}\rightarrow 0$. 
Thus we obtain 
the Ward identity for heat current vertex, 
\begin{equation}
\sum_{\mu=0}^3 k_\mu \Gamma_\mu^Q(p+k,p) = 
p_0 G^{-1}(p+k) - (p_0+k_0) G^{-1}(p), \label{WI-Q} 
\end{equation}
where 
$\Lambda_\mu^Q(p+k,p) = i G(p+k) \cdot \Gamma_\mu^Q(p+k,p) \cdot i G(p)$. 

\section{Ward identities for Cooper pairs}

First we introduce an anisotropic Cooper pair~\cite{LV} 
represented by its center-of-mass coordinate ${\vec R}$ 
\begin{equation}
\Psi({\vec R}) = 
\int d {\vec r} \chi_l({\vec r}) 
\psi_\downarrow({\vec r_1}) \psi_\uparrow({\vec r_2}), 
\end{equation}
where 
\begin{equation}
\chi_l({\vec r}) = 
\sum_{\vec p} e^{i{\vec p}\cdot{\vec r}} \chi_l({\vec p}),\ \ \ 
\psi_\downarrow({\vec r_1}) = 
\sum_{\vec p_1} e^{i{\vec p_1}\cdot{\vec r_1}} b_{\vec p_1},\ \ \ 
\psi_\uparrow({\vec r_2}) = 
\sum_{\vec p_2} e^{i{\vec p_2}\cdot{\vec r_2}} a_{\vec p_2}, 
\end{equation}
with the relative coordinate ${\vec r}$ 
and thus 
${\vec r_1}={\vec R}+{\vec r}/2$ and ${\vec r_2}={\vec R}-{\vec r}/2$. 
Performing the integral in terms of the relative coordinate ${\vec r}$ 
we obtain 
\begin{equation}
\Psi({\vec R}) = 
\sum_{\vec q} e^{i{\vec p}\cdot{\vec R}} P_{\vec q}, 
\end{equation}
with 
\begin{equation}
P_{\vec q} = 
\sum_{\vec p} \chi_l({\vec p}) 
b_{{{\vec q} \over 2}-{\vec p}} a_{{{\vec q} \over 2}+{\vec p}}, \ \ \ \ \ 
P_{\vec q}^\dag = 
\sum_{\vec p} \chi_l({\vec p}) 
a_{{{\vec q} \over 2}+{\vec p}}^\dag b_{{{\vec q} \over 2}-{\vec p}}^\dag. 
\end{equation}
The form factor $\chi_l({\vec p})$ has been introduced 
through the interaction Hamiltonian $H_{\rm int}$ as 
\begin{equation}
H_{\rm int} = 
\sum_{\vec q} \sum_{\vec p} \sum_{\vec p'} 
V({\vec p},{\vec p'}) 
a_{{{\vec q} \over 2}+{\vec p}}^\dag b_{{{\vec q} \over 2}-{\vec p}}^\dag 
b_{{{\vec q} \over 2}-{\vec p'}} a_{{{\vec q} \over 2}+{\vec p'}}, 
\end{equation}
where 
\begin{equation}
V({\vec p},{\vec p'}) = - 
g_l \chi_l({\vec p}) \chi_l({\vec p'}), 
\end{equation}
with 
$g_l$ being the strength of the attractive interaction 
of BCS-type 
for $l$-wave pairing. 
Here we consider only spin-singlet pair for simplicity. 

In the case of electric current 
the three-point function $M_\mu^e$ for Cooper pairs 
satisfies the relation 
\begin{align}
\sum_{\mu=0}^3 i k_\mu M_\mu^e(q,q-k) = 
\int d & (x_0-y_0) e^{-iq_0(x_0-y_0)} \int d (z_0-x_0) e^{-ik_0(z_0-x_0)} 
\nonumber \\ 
\times \Bigl(
& \langle T 
\{ [j_{\vec k}^e(x_0), P_{{\vec q}-{\vec k}}(x_0)] P_{\vec q}^\dag(y_0) \} 
\rangle \delta(z_0-x_0) \nonumber \\ 
+ 
& \langle T 
\{ P_{{\vec q}-{\vec k}}(x_0) [j_{\vec k}^e(y_0), P_{\vec q}^\dag(y_0)] \} 
\rangle \delta(z_0-y_0) \Bigr). \label{divLambda-e-Cooper} 
\end{align}
The equal time commutation relations are calculated as 
\begin{equation}
[ j_{\vec k}^e, P_{{\vec q}-{\vec k}} ] = - e 
\sum_{\vec p} \chi_l({\vec p}) 
\Bigl(
b_{{{\vec q} \over 2}-({\vec p}+{{\vec k} \over 2})} 
a_{{{\vec q} \over 2}+({\vec p}+{{\vec k} \over 2})} 
+ 
b_{{{\vec q} \over 2}-({\vec p}-{{\vec k} \over 2})} 
a_{{{\vec q} \over 2}+({\vec p}-{{\vec k} \over 2})} 
\Bigr), 
\end{equation}
and 
\begin{equation}
[ j_{\vec k}^e, P_{\vec q}^\dag ] = e 
\sum_{\vec p} \chi_l({\vec p}) 
\Bigl(
a_{{{\vec q}-{\vec k} \over 2}+({\vec p}+{{\vec k} \over 2})}^\dag 
b_{{{\vec q}-{\vec k} \over 2}-({\vec p}+{{\vec k} \over 2})}^\dag 
+ 
a_{{{\vec q}-{\vec k} \over 2}+({\vec p}-{{\vec k} \over 2})}^\dag 
b_{{{\vec q}-{\vec k} \over 2}-({\vec p}-{{\vec k} \over 2})}^\dag 
\Bigr). 
\end{equation}
If the pairing symmetry is isotropic $\chi_l({\vec p})=1$, 
these commutation relations picks up 
the charge of the Cooper pair $2e$ as 
\begin{equation}
[ j_{\vec k}^e, P_{{\vec q}-{\vec k}} ] = - 2 e 
\sum_{\vec p'} 
b_{{{\vec q} \over 2}-{\vec p'}} 
a_{{{\vec q} \over 2}+{\vec p'}} 
= - 2 e P_{\vec q}, 
\end{equation}
and 
\begin{equation}
[ j_{\vec k}^e, P_{\vec q}^\dag ] = 2 e 
\sum_{\vec p'} 
a_{{{\vec q}-{\vec k} \over 2}+{\vec p'}}^\dag 
b_{{{\vec q}-{\vec k} \over 2}-{\vec p'}}^\dag 
= 2 e P_{{\vec q}-{\vec k}}^\dag, 
\end{equation}
where we have shifted the variable of the summation. 
Even for anisotropic Cooper pairs 
the same commutation relations can be used 
in the limit of vanishing external momentum, ${\vec k}\rightarrow 0$, 
where 
the shift in the argument of the form factor 
$\chi_l({\vec p'}\pm{{\vec k}\over 2})$ 
is negligible. 
Thus introducing the Cooper pair propagator\footnote{
The fluctuation propagator ($T>T_c$) for anisotropic Cooper pairs 
is discussed, for example, in ref.~\cite{ERS}.} $D(q)$ with four-momentum as 
\begin{equation}
D(q) = 
\int d (x_0-y_0) e^{-iq_0(x_0-y_0)} D_{\vec q}(x_0-y_0), 
\end{equation}
with 
\begin{equation}
D_{\vec q}(x_0-y_0) = 
- i \langle T \{ P_{\vec q}(x_0) P_{\vec q}^\dag(y_0) \} \rangle, 
\end{equation}
we obtain the Ward identity 
\begin{equation}
\sum_{\mu=0}^3 k_\mu \Delta_\mu^e(q+k,q) = 
2e D^{-1}(q) - 2e D^{-1}(q+k), \label{WIC-e}
\end{equation}
for electric current vertex 
where 
$M_\mu^e(q+k,q) = i D(q+k) \cdot \Delta_\mu^e(q+k,q) \cdot i D(q)$. 

In the case of heat current 
the three-point function $M_\mu^Q$ for Cooper pairs 
satisfies the relation 
\begin{align}
\sum_{\mu=0}^3 i k_\mu M_\mu^Q(q,q-k) = 
\int d & (x_0-y_0) e^{-iq_0(x_0-y_0)} \int d (z_0-x_0) e^{-ik_0(z_0-x_0)} 
\nonumber \\ 
\times \Bigl(
& \langle T 
\{ [j_{\vec k}^Q(x_0), P_{{\vec q}-{\vec k}}(x_0)] P_{\vec q}^\dag(y_0) \} 
\rangle \delta(z_0-x_0) \nonumber \\ 
+ 
& \langle T 
\{ P_{{\vec q}-{\vec k}}(x_0) [j_{\vec k}^Q(y_0), P_{\vec q}^\dag(y_0)] \} 
\rangle \delta(z_0-y_0) \Bigr). \label{divLambda-Q-Cooper} 
\end{align}
The equal time commutation relations for anisotropic Cooper pairs 
are estimated as 
\begin{equation}
[ j_{\vec k}^Q, P_{{\vec q}-{\vec k}} ] \Rightarrow 
[ H, P_{\vec q} ], \ \ \ \ \ 
[ j_{\vec k}^Q, P_{\vec q}^\dag ] \Rightarrow 
[ H, P_{{\vec q}-{\vec k}}^\dag ], 
\end{equation}
in the limit of vanishing external momentum, ${\vec k}\rightarrow 0$. 
Thus 
we obtain the Ward identity 
\begin{equation}
\sum_{\mu=0}^3 k_\mu \Delta_\mu^Q(q+k,q) = 
q_0 D^{-1}(q+k) - (q_0+k_0) D^{-1}(q), \label{WIC-Q}
\end{equation}
for heat current vertex 
where 
$M_\mu^Q(q+k,q) = i D(q+k) \cdot \Delta_\mu^Q(q+k,q) \cdot i D(q)$. 

\section{Conclusion}

We have derived the Ward identities for anisotropic Cooper pairs. 
The effect of the anisotropy can be taken into account as the form factor 
so that the resulting Ward identities are the same 
as those for isotropic pairs. 


\end{document}